\documentclass[aps,prl,twocolumn,amsmath,amssymb, superscriptaddress]{revtex4-1}
\usepackage[utf8]{inputenc}
\usepackage{graphicx}
\usepackage{dcolumn}
\usepackage{bm}
\usepackage[colorlinks,linkcolor=blue,anchorcolor=blue,citecolor=blue,urlcolor=blue,filecolor=blue,menucolor=blue,runcolor=blue]{hyperref}
\usepackage{ulem}
\usepackage{siunitx}
\usepackage{xr-hyper}   
\usepackage{hyperref}
\makeatletter
\newcommand*{\addFileDependency}[1]{%
  \typeout{(#1)}\@addtofilelist{#1}%
  \IfFileExists{#1}{}{\typeout{No file #1.}}}
\makeatother
\newcommand*{\myexternaldocument}[2][]{%
  \externaldocument[#1]{#2}%
  \addFileDependency{#2.tex}%
  \addFileDependency{#2.aux}}

\myexternaldocument[S-]{supplementary}

\makeatletter
\def\@email@name{Email address}

\@ifundefined{textcolor}{}
{%
	\definecolor{BLACK}{gray}{0}
	\definecolor{WHITE}{gray}{1}
	\definecolor{RED}{rgb}{1,0,0}
	\definecolor{GREEN}{rgb}{0,1,0}
	\definecolor{BLUE}{rgb}{0,0,1}
	\definecolor{CYAN}{cmyk}{1,0,0,0}
	\definecolor{MAGENTA}{cmyk}{0,1,0,0}
	\definecolor{YELLOW}{cmyk}{0,0,1,0}
}

\usepackage{color}

\makeatother
\usepackage[english]{babel}

\begin{document}

\title{Current-driven resistive switching in high-field heavy fermion metal UTe$_2$}

\author{Ling Zhang}
\email{ling.zhang@mpsd.mpg.de}
\affiliation{Max Planck Institute for the Structure and Dynamics of Matter, 22761 Hamburg, Germany}

\author{Chunyu (Mark) Guo}
\email{chunyu.guo@mpsd.mpg.de}
\affiliation{Max Planck Institute for the Structure and Dynamics of Matter, 22761 Hamburg, Germany}

\author{David Graf}
\affiliation{National High Magnetic Field Laboratory, Tallahassee, FL 32310, USA}

\author{Carsten Putzke}
\affiliation{Max Planck Institute for the Structure and Dynamics of Matter, 22761 Hamburg, Germany}

\author{Mitchell M. Bordelon}
\affiliation{Los Alamos National Laboratory, Los Alamos, NM 87545, USA}

\author{Eric D. Bauer}
\affiliation{Los Alamos National Laboratory, Los Alamos, NM 87545, USA}

\author{Sean M. Thomas}
\affiliation{Los Alamos National Laboratory, Los Alamos, NM 87545, USA}
\affiliation{Department of Physics, Colorado State University, Fort Collins, Colorado 80523, USA}

\author{Filip Ronning}
\affiliation{Los Alamos National Laboratory, Los Alamos, NM 87545, USA}

\author{Priscila F. S. Rosa}
\affiliation{Los Alamos National Laboratory, Los Alamos, NM 87545, USA}
\affiliation{Department of Physics, Colorado State University, Fort Collins, Colorado 80523, USA}

\author{Philip J. W. Moll}
\email{philip.moll@mpsd.mpg.de}
\affiliation{Max Planck Institute for the Structure and Dynamics of Matter, 22761 Hamburg, Germany}

\begin{abstract}
In UTe$_2$, superconductivity persists up to 70 T for fields tilted by roughly 20–40° from the $b$ axis toward $c$. Its normal state is a field-polarized (FP) state, which emerges above a first-order metamagnetic transition when the field component along $b$ exceeds $\sim$35 T. Despite being the parent of this high-field superconductivity, the FP state remains poorly understood, including whether it shares underlying physics with the low-field paramagnetic (PM) state from which it is abruptly separated. Using microstructures fabricated by focused ion beam, we reach current densities of 30–60 kA/cm$^2$ and observe a current-induced high-resistivity state in both the PM and FP regimes over a wide range of field angles in the $bc$ plane. Above a critical current density of order 10 kA/cm$^2$, the resistivity sharply rises to values comparable to those near the critical end point of the metamagnetic transition. The critical current density increases with temperature, opposite to the trend expected from Joule heating, ruling out a thermal origin. The emergence of the same instability on both sides of the metamagnetic transition reveals an unexpected commonality between the FP and PM states. More broadly, the observation of a current-induced transition in a heavy-fermion metal exposes the multitude of nearby electronic states in correlated electron systems and shows that electric current, like magnetic field and pressure, can switch between them.
\end{abstract}

\maketitle

\section{Introduction}

UTe$_2$ is one of the most intriguing superconductors in contemporary condensed matter physics, owing in large part to its unusual magnetic-field-temperature phase diagram. Despite its relatively modest superconducting transition temperature, $T_c \sim 2.1$ K, superconductivity persists to exceptionally high magnetic fields, with upper critical fields reaching approximately 70 T in certain field orientations—far above conventional Pauli-limiting estimates\cite{aokiUnconventionalSuperconductivityHeavy2019,ranNearlyFerromagneticSpintriplet2019, knebelFieldReentrantSuperconductivityClose2019, lewinHighfieldSuperconductingHalo2025, ranExtremeMagneticFieldboosted2019, wuEnhancedTripletSuperconductivity2024}. These properties have made UTe$_2$ a prominent candidate for spin-triplet superconductivity. Ferromagnetic fluctuations, by analogy with other uranium-based superconductors\cite{aokiReviewUbasedFerromagnetic2019} such as UCoGe and URhGe, were initially considered a possible pairing mechanism\cite{ranNearlyFerromagneticSpintriplet2019}. Subsequent experiments have observed antiferromagnetic fluctuations at ambient pressure\cite{duanIncommensurateSpinFluctuations2020} and antiferromagnetic order at high pressure\cite{thomasEvidencePressureinducedAntiferromagnetic2020,knafoIncommensurateAntiferromagnetismUTe22025}, challenging this hypothesis. 

The overall close connection between magnetism and superconductivity is particularly evident in the high-field phase diagram. The high-field phase diagram contains three different superconducting states (SC1-3) existing within two magnetically distinct regimes separated by a strongly first-order metamagnetic transition (Fig. 1). This boundary sharply truncates the field-reinforced superconducting phase SC2 near 36 T for $H\parallel b$, while superconductivity reappears at still higher fields in a separate region, SC3. Whether SC2 and SC3 share a common pairing mechanism and order-parameter symmetry remains an open question. Their markedly different sensitivity to crystalline disorder supports the possibility of distinct superconducting states as SC3 survives disorder levels that fully suppress SC2 and even zero-field superconductivity\cite{frankOrphanHighField2024}. Conversely, the proximity of SC2 and SC3 in the field–angle phase diagram, and their apparent meeting near a critical angle\cite{aokiHighFieldSuperconducting2024}, suggest that both are strongly influenced by the same underlying magnetic energy scales. Establishing whether this connection reflects a common pairing mechanism or merely a shared magnetic environment requires a better understanding of the normal states from which the two superconducting phases emerge.

The magnetic character of these normal states remains incompletely understood, in part because the extreme fields required to access them severely constrain microscopic probes. The metamagnetic transition is accompanied by a large jump in the longitudinal magnetization\cite{miyakeMetamagneticTransitionHeavy2019}, motivating the designation “field-polarized” (FP) for the high-field state. The lower-field regime is commonly termed “paramagnetic” (PM), reflecting the absence of any magnetic order at zero field and the lack of a known intervening magnetic phase transition. These labels are useful phenomenologically, but they should not be taken to imply that the FP state consists of saturated moments or that the PM state is magnetically uncorrelated. Instead, both are highly complex magnetic states on an equal footing.

Here, we experimentally discover a surprising instability of both the PM and FP phases under strong dc current drive. Both exhibit a sharply defined threshold current, above which the transport response changes to a common high-current state. The similarity of these responses provides a new experimental perspective on the direct connection between the two magnetic regimes and places constraints on the mechanisms underlying their field-dependent transport. The instability of the magnetic states under transport currents is a rare response that calls for theoretical efforts towards a microscopic understanding.

\section{Current-driven transport switching in high-field paramagnetic phases}

To investigate non-linear transport in UTe$_2$ under extreme current densities, the cross-sectional area of the current channel must be substantially reduced compared to bulk single crystals. We therefore employ focused ion beam (FIB) micro-fabrication to pattern micron-scale transport devices\cite{mollFocusedIonBeam2018}. A total current of $2\ \mathrm{mA}$ translates to a current density of up to $\sim 70\ \mathrm{kA/cm}^2$ in the microbars, thereby limiting the heating in the wires and contacts. Details of the fabrication procedure are described in our previous work\cite{zhangDimensionalityReinforcedSuperconductivity2025}. Electrical transport measurements, including resistivity and current-voltage (IV) characteristics, are primarily performed in a DC configuration (see Methods), with a pulsed IV method applied where indicated to mitigate Joule heating at high current densities. 

Before examining current-driven transport, we establish that the FIB-fabricated microstructures reproduce the known high-field phase structure of UTe$_2$ in the low-current limit ($J<0.3\ \mathrm{kA/cm}^2$), well within the linear-transport regime. Field sweeps of $\rho_a$ at $T=0.35~\mathrm{K}$ for magnetic fields applied along different directions between the $b$ and $c$ axes exhibit the known metamagnetic transition as well as the field reinforced SC states SC2 and SC3 (Fig. 1b). The vertical nature of the phase boundaries of SC2 and SC3 is best probed by continuously rotating the sample within the $bc$ plane at fixed field (Fig. 1c). In this protocol, phase boundaries are crossed by changing the field angle rather than the field magnitude. The abrupt jump in $\rho_a(\theta)$ at $\theta_c^{\mathrm{MM}}\approx35^\circ$ marks the metamagnetic transition, which broadens into a peak near $10~\mathrm{K}$, consistent with the established critical end point of this transition \cite{miyakeMetamagneticTransitionHeavy2019,knafoComparisonTwoSuperconducting2021}. As the angle decreases further, the system enters the SC3 phase. At each temperature, its low-angle boundary occurs where the upper critical field falls to $44~\mathrm{T}$, the fixed field value for the angle sweeps. The large geometric factor of the microstructured devices provides a high sensitivity to resistivity, which is particularly valuable in SC3: in a meander-shaped device, the resistivity in SC3 is bounded above by $0.14~\mathrm{n}\Omega\,\mathrm{cm}$, more than six orders of magnitude below that of the adjacent normal state, supporting its assignment as a superconducting state rather than an exceptionally good metal (see Supplementary Material).

This suite of experimental data, including the angle-dependent phase diagram, the magnetoresistance traces and their angle dependence, agrees quantitatively with bulk single-crystal measurements\cite{helmFieldinducedCompensationMagnetic2024, lewinHighfieldSuperconductingHalo2025, wuSuperconductingCriticalTemperature2025}. The unchanged high crystalline quality of the microstructures is further evidenced by the sharp resistive transition at a high $T_c\sim2.1~\mathrm{K}$, in light of the known disorder sensitivity of SC1, as well as the clear observation of Shubnikov-de Haas oscillations in the microstructures\cite{zhangDimensionalityReinforcedSuperconductivity2025}. All the data presented are quantitatively reproduced in several other samples (see Supplementary Table S1 for the samples used in each figure).

As an initial characterization of the current-driven transport response in UTe$_2$, we measured $\rho_a(\theta)$ at $T=0.3~\mathrm{K}$ using the same angle-sweeping protocol as in the low-current map, but with the applied probing DC current density increased up to a maximum of $J=34~\mathrm{kA/cm^2}$ (Fig. 2a). Indeed, the transport is strongly non-linear over the entire angle range, governed by distinct physical processes, including thermal effects, as we demonstrate below. At low current densities, the zero-resistance SC3 region is progressively suppressed and disappears above $J=17~\mathrm{kA/cm^2}$, followed by a pronounced resistivity peak near the MM transition at $J\geq18~\mathrm{kA/cm^2}$. As the current density increases further, the sharp resistive peak broadens into an extended high-resistivity state. Close to the $b$ axis, the resistivity telegraphs irregularly between low- and high-resistivity levels, indicating a critical bistability. Notably, at high current densities this state also extends to high angles beyond the MM critical angle identified at low current densities, demonstrating that it is accessible at field angles that yield a PM ground state at zero bias. This suggests the current-driven high-resistivity state as a common instability of both high-field states, PM and FP alike, despite their distinct low-current transport properties. 

Joule heating accompanies finite-resistance transport, and it is key to identify the role of self-heating in its non-linearity. Although submersion of our samples in liquid $^3$He improves thermal anchoring, self-heating may in principle induce thermal bistability in conductors and thereby mimic a transition-like response\cite{gurevichSelfheatingNormalMetals1987}. While self-heating is present, the angular dependence of $\rho _a$ does not indicate a thermal origin of the transition. As the field is rotated towards the critical angle from higher angles, $\rho_a$ remains essentially constant before increasing abruptly at the transition. Since the applied current is fixed, the dissipated power is approximately unchanged throughout this rotation. The transition therefore does not result from a progressive increase of heating, but is triggered by the field angle under essentially constant thermal conditions.

The non-linear $E$-$J$ characteristics provide clear insight into the threshold behavior of the system (Fig. 2b). In all $E$-$J$ measurements, we use a pulsed-current configuration to minimize cumulative Joule heating (pulse width set to 1 ms, delay time between pulses set to 200 ms; see Methods for further details). At the base temperature $T=0.3~\mathrm{K}$, the $E$-$J$ curve in the PM state at $\theta=54^\circ$ remains linear up to remarkably high current densities above $10~\mathrm{kA/cm^{2}}$. At a sharp threshold current of $J_{\mathrm{c}}^*=36~\mathrm{kA/cm^2}$, the sample then abruptly transitions into the high-resistance state. 

The temperature dependence of the threshold self-evidently demonstrates that heating does not trigger this transition. Increasing the bath temperature from $0.3~\mathrm{K}$ to around $1~\mathrm{K}$ increases $J_{\mathrm{c}}^*$, as confirmed in multiple samples, hence the transition requires substantially less power at lower temperatures. In a heating-driven scenario, the opposite is expected: a higher bath temperature reduces the additional heating needed to reach the transition, and thus lowers the threshold power (see Supplementary Material). We therefore identify the high-resistance state as an intrinsic, current-driven instability of the high-field paramagnetic phase.

\section{A current-defined instability landscape of the high-field states}

Having identified a current-induced transition into a high-resistivity state, we next examine how this instability evolves across the high-field phase space (Fig. 3). Since the system lies in the superconducting region at some angles (e.g., $\theta=31^\circ$, Fig. 3c), two critical phenomena appear in one trace: first, the superconducting critical current at low bias, and second, the high-resistance transition in the normal state at high bias. To display both phenomena on the same scale, we turn here to $\rho$–$J$ plots, showing resistivity instead of electric field; Fig. 3d replots the $\theta=54^\circ$ data of Fig. 2b in this representation. At base temperature, $T=0.3~\mathrm{K}$, the high resistance transition occurs at all measured angles, consistent with the broad plateau in the angular sweep in Fig. 2a. The three distinct equilibrium states at this temperature (FP at $\theta=12^\circ$ and $22^\circ$, Fig. 3a,b; SC3 at $\theta=31^\circ$, Fig. 3c; PM at $\theta=54^\circ$, Fig. 3d) present a remarkably consistent non-linear signature: the current-induced transition is similarly sharp at all angles and emerges suddenly from a flat resistive background. By contrast, the high-resistivity state reached above $J_c^*$ shows a pronounced increase of resistance with increasing current at all angles. It is notable that this trend continues deep into the high-resistivity state: even at the highest drive currents, the resistance at base temperature remains clearly higher than that at elevated temperatures. Upon warming, $J_c^*$ initially increases at all angles, as at $\theta=54^\circ$, indicating that this anomalous temperature evolution persists under substantial changes in field orientation.  At even higher temperatures, $T>1.75~\mathrm{K}$, the sharp transition gives way to a gradual resistivity enhancement, most clearly seen at $\theta=12^\circ$ and $31^\circ$, with an onset current density lower than that at $T=1~\mathrm{K}$.

The behavior within SC3 clearly stands out (Fig. 3c). At low currents, its zero resistance state terminates at a superconducting critical current. A double-stepped transition is observed which we assign to separate transitions in the transport bar and the contact regions due to current inhomogeneities. At $\theta=31^\circ$ and $T=0.3~\mathrm{K}$, a superconducting critical current of $J_c\approx 10 ~\mathrm{kA/cm^2}$ is measured. This is compatible with typical values found in heavy fermion superconductors\cite{wuchnerMagneticPropertiesHeavyfermion1993, majumdarAnisotropicLowfieldBehavior2003, polliniFluxDynamicsLowfield1993, wulferdingLocalCharacterizationHeavyfermion2020}. This value, however, is notable given that it is measured in a single crystal at 44 T in a high-density vortex state. In optimized cuprate conductors for high-field applications, $500~\mathrm{kA/cm^2}$ is typical for the critical current density of the superconducting material at $30~\mathrm{T}$ and 4.2 K\cite{jiangHighPerformanceBi2212Wires2019}. Clearly, our geometry and voltage sensitivity do not provide the rigorous quantification as used in industry; still, the measurement provides a reference scale: a single crystal of a 2.1 K heavy-fermion superconductor reaches approximately 2\% of the critical current density of an engineered high-performance conductor with a substantially higher $T_c\sim 80~\mathrm{K}$. Moreover, this critical current density is significantly higher than those reported for the SC1 and SC2 phases, which are remarkably small, $\sim 10^{-3}~\mathrm{kA/cm^2}$\cite{wuElectricallyControllableSuperconducting2026, zhangDimensionalityReinforcedSuperconductivity2025}. This demonstrates not only that the order parameter of SC3 is more robust to high fields, but also that vortices at high density can be superbly pinned in this phase, suggesting an additional pinning mechanism for the SC3 phase. 

A separate qualitative distinction at this angle is the stronger current dependence of resistance in the low-resistance state, above the critical current of the superconducting state. In Fig. 3c, the curves at all temperatures below 1.5 K collapse onto one approximately linear $\rho$–$J$ curve, in contrast to the constant resistivity at other angles. A thermal origin of this slope change appears unlikely, given that no obvious nonlinear behavior is observed at $22^\circ$ at similar resistivity and current, and hence similar heating power. The trend of increasing critical current of the high-resistance transition with temperature remains unchanged, despite the qualitatively different behavior of the low-resistance state and the growing dissipation with increasing current. This points to an instability of the non-linear conduction as the sample is rotated towards the critical angle of the MM transition.

The evolution of the critical current with angle further exposes the critical response of the system to proximity to the MM line (Fig. 4a). As the resistive transitions broaden into a crossover from low to high resistivity, we use a universal voltage threshold to estimate the critical current, analogous to the criterion used for a superconducting critical current (see supplementary). In this wide temperature range, the $J_c^*(\theta)$ curves consistently exhibit a pronounced dip near the MM critical angle, $\theta_c^{\mathrm{MM}}\approx34^\circ$. The FP state appears to remain rather susceptible to a current-induced instability, with a weaker and eventually saturating angle dependence as $H\parallel b$ ($\theta=0^\circ$) is approached. On the PM side, however, $J_\mathrm{c}^*$ rises more steeply as the field is rotated away from the MM transition, indicating that the high-resistivity state becomes rapidly harder to reach.

These observations can be summarized by extending the phase diagram to include current density $J$ as an additional control parameter (Fig. 4b). At low temperature, increasing $J$ drives the system across a first-order-like transition (purple plane) into a strongly driven state. This high-current state closely resembles the state above the metamagnetic critical endpoint $T_{\mathrm{CEP}}$: both have a similar field-angle dependence and comparable resistance magnitude. Moreover, on heating from the high-current state into the high-temperature state (orange dashed path in Fig. 4b), we detect no signature of an additional transition, indicating that the two are continuously connected. Since the FP and PM states are themselves connected by a crossover above $T_{\mathrm{CEP}}$, all four regions are consistent with sharing the same symmetry. A first-order boundary separating states of identical symmetry, like the liquid–gas line, need not extend indefinitely and can instead terminate at a line of critical points. We therefore expect the current-driven first-order plane to also end at a critical line with increasing temperature, and this critical line likely merges with the metamagnetic critical endpoint in the zero-current limit.

\section{Discussion}

These observations taken together demonstrate a step-like transition in UTe$_2$ driven by large current densities, suggestive of a breakdown of the magnetic texture formed in the FP and PM phases below 10K. The microscopics of its superconducting ground state, such as the order parameter symmetry and pairing mechanism, remain debated and much less is known about the reentrant phases at high magnetic fields. Both the limitations of experimental techniques, in particular local spectroscopies, and the difficulty of treating magnetic fields non-perturbatively in the interacting many-body problem make this one of the most complex puzzles in correlated matter physics. The sensitivity to applied current adds yet another experimental facet to help refine future theoretical models, yet a microscopic understanding of this phenomenon is far out of reach at the moment.

However, one can draw conclusions from these observations and identify key ingredients. First, the heavy-fermion nature of UTe$_2$ amplifies the relative effects of static electric fields. Within a Boltzmann framework, a larger shift of the Fermi surface is needed to sustain a certain current density owing to the low Fermi velocity of the heavy quasiparticles. For usual metals at the relevant current scale $J=10^4~\mathrm{A/cm^2}$, a quick estimate yields $\delta k/k_{\mathrm{F}} = m^*J /ne\hbar k_{\mathrm{F}}  \sim 5\times10^{-9}$ using copper as a typical example ($m^*\sim m_e$, $n\sim 10^{23} ~\mathrm{cm^{-3}},k_{\mathrm{F}}\sim 1.3$ \AA{}$^{-1}$). This shift is negligible, reflecting the fact that the electronic properties at the Fermi surface govern metals and that they usually are linear conductors. UTe$_2$ hosts massive electrons ($m^*\sim 40 m_e$) at lower density ($n\sim 10^{22}~\mathrm{cm^{-3}}, k_{\mathrm{F}}\sim 0.6$ \AA{}$^{-1}$), hence fewer and slower electrons need a larger momentum shift to carry this current, $\delta k/k_{\mathrm{F}}\sim 4\times10^{-6}$. Although this displacement is approximately 800 times larger than in the copper example, it remains a small fraction of the Fermi momentum, helping to explain why nonlinear current-driven phenomena are uncommon in heavy-fermion materials despite a few exotic exceptions\cite{mollFieldinducedDensityWave2015, ronningElectronicInplaneSymmetry2017}. However, this momentum imbalance may impact the dynamic structure forming in the magnetic fluctuation spectrum, which sets the high-field physics of the FP and PM states.

A second mechanism could arise from modifications of the Kondo effect. This is well studied in gapped Kondo systems, such as the Kondo breakdown in quantum dots\cite{roschKondoEffectQuantum2001}; however, such physics is not straightforwardly translated to gapless systems that screen electric fields. A more feasible picture therefore would be related to the coherent heavy Fermi liquid ground state. Although current evidence does not conclusively establish a heavy-Fermi-liquid state in UTe$_2$ at high magnetic fields, we use this picture as one of the possible frameworks for interpreting the current-driven response. As shown in Fig.4c, Kondo coherence can be weakened when the system deviates further from equilibrium under strong current drive. To obtain a finite energy scale, we make phenomenological estimate by comparing the energy acquired by a heavy electron over the transport mean free path to the Kondo-coherence energy $ eE\ell_{\mathrm{tr}}\sim k_{\mathrm{B}}T_{\mathrm{coh}}$. In this picture, the coherent state is not destabilized directly via momentum imbalance, but rather via inelastic scattering of accelerated electrons. This comparison yields an electric field of order $10^3~\mathrm{V/m}$ for a Kondo coherence temperature $\sim 10~\mathrm{K}$ in UTe$_2$, corresponding to a current density of $J\sim 100 ~\mathrm{kA/cm^2}$, only one order of magnitude above the critical electric current observed experimentally (see discussion in supplementary). A non-equilibrium distribution of hot electrons may scatter off the incipient magnetic order, delivering the energy required to destabilize it. This offers an attractive direction to explain the growing critical current with increasing temperature, because thermal broadening increases the electron-electron scattering rate as $\Gamma\sim (T/T_F)^{2}$, limiting the acceleration of individual quasiparticles without relaxing the total current. 

A third natural hypothesis would be an excitation of a static order parameter above a threshold drive, such as depinning of density waves, domain walls, or magnetic textures such as skyrmions\cite{grunerDynamicsChargedensityWaves1988,grunerDynamicsSpindensityWaves1994,yamaguchiRealSpaceObservationCurrentDriven2004,jonietzSpinTransferTorques2010}. Importantly, no such order is known in zero field and no evidence for such a field-induced phase has been reported. The MM line is the only high-field transition in the normal state, yet it is clearly first order and a continuous trajectory around its critical end point equally connects the FP state to zero field without crossing a known phase transition. This scenario hinges on excitations of a thus far undetected order parameter, associated either with zero-field order that leaves no thermodynamic anomaly or with a dark phase transition in field that escapes detection by the limited probes available at high fields. 

Alternatively, in the absence of static order, a current-driven modification of the spin-fluctuation spectrum is a natural candidate. Hints that this spectrum hosts structure on low energy scales come from ripple-like features recently observed at the metamagnetic transition for selected field orientations\cite{wuMetamagneticRipplesUTe22025}, which were proposed to reflect an additional low-energy scale or metastability near the phase boundary. As these features are not accompanied by any additional thermodynamic phase transition, they point to nearly degenerate states of the same symmetry rather than to a hidden order parameter, in line with the continuous connection between the high-current and high-temperature states found here (Fig.~4b). An intriguing parallel is offered by the antiferromagnetic phase stabilized under pressure, which can be viewed as lying on the same side of the metamagnetic boundary as the FP state once that boundary is followed from the field axis to the pressure axis\cite{ranExpansionHighFieldboosted2021}. There, neutron scattering finds a change of the ordering wave vector at temperatures as low as 1--2~K\cite{knafoIncommensurateAntiferromagnetismUTe22025}. It is tempting to speculate that the fluctuating counterpart of this order at ambient pressure and high magnetic fields retains similarly low energy scales, leaving it susceptible to modest perturbations such as an applied current. Whether such fluctuating magnetic objects can be depinned by currents, in analogy to static textures such as skyrmions, remains an open question. That they couple to the charge carriers is evident from the reduction of resistance below 10~K at high magnetic fields (Fig.~1c).

Independent of the microscopic mechanism, the close connection between the current-induced and metamagnetic transitions constrains how the high-current state differs from the low-current one. Across the metamagnetic transition, transport measurements indicate a reconstruction of the Fermi surface\cite{niuEvidenceFermiSurface2020}, which may also occur across the current-induced transition. Since the high-current state resembles the state above $T_{\mathrm{CEP}}$, such a reconstruction would likely bring the electronic structure towards that of the high-temperature regime. More generally, a change of electronic state, whether in the Fermi surface itself or in the scattering of carriers off magnetic fluctuations, offers a natural account of the step in resistance to a magnitude comparable to that above $T_{\mathrm{CEP}}$. By contrast, the rigid displacement of the Fermi surface estimated above, $\delta k/k_{\mathrm{F}}\sim 4\times10^{-6}$, is far too small to change transport appreciably. By the same analogy, the transition should also be reflected in the magnetization, which takes the role of the density in the liquid--gas picture. Below $T_{\mathrm{CEP}}$, magnetization curves $M(T)$ measured just below and just above the metamagnetic field split from a common high-temperature curve into separate PM and FP branches\cite{miyakeMetamagneticTransitionHeavy2019}. Being continuously connected to the high-temperature regime, the current-induced state would probably correspond to the continuation of that common branch to low temperatures.

Our inability to describe complex strongly interacting electron systems with multiple comparable energy scales, such as the Zeeman energy, is compounded by the technical challenges of working with high-field magnets at the limits of current materials engineering, low-temperatures and now also at high bias current densities. At the same time, the accidental discoveries of the rich structures formed by strong interactions provide a glimpse of what material behaviors will become possible and controllable in the future once such systems are understood. For now, our experimental results demonstrate that even the normal state of UTe$_2$ is far from an ordinary metal. Following the estimates above, light as well as heavy fermion metals should not generically show current driven instabilities, in line with experimental observations of linear transport in them. The remarkable fact that such a transition exists here adds another hint that new physics may govern its high field state.

\section{Methods}

\subsection{Crystal synthesis and characterization}
Single crystals of UTe$_2$ were grown by the molten salt flux (MSF) method using an equimolar NaCl + KCl mixture\cite{sakaiSingleCrystalGrowth2022}, and characterized by single-crystal x-ray diffraction and specific heat measurements as described previously\cite{zhangDimensionalityReinforcedSuperconductivity2025}. Specific heat confirmed a single superconducting transition at $T_c = 2.1~\mathrm{K}$.

\subsection{Microstructure fabrication}
Microstructures were fabricated following the focused ion beam (FIB) procedure described in Ref.~\onlinecite{zhangDimensionalityReinforcedSuperconductivity2025}. In brief, 6 lamellae were cut from an MSF-grown crystal, transferred onto sapphire substrates with FIB-deposited Pt. Each lamella was Ar-ion etched to remove the surface oxide layer and then, without breaking vacuum, coated with Au by DC magnetron sputtering (samples S1,S2,S4,S5,S6) or high-power impulse magnetron sputtering (HiPIMS; sample S3) to form electrical contacts. All other steps were identical. 

\subsection{High-field transport measurements}
High-field transport measurements were performed in the 45 T hybrid magnet at the National High Magnetic Field Laboratory. For field- and angle-sweep measurements, the voltage was measured in one of two ways. In the first, Stanford Research Systems SR86x lock-in amplifiers read the signal, with the AC excitation current supplied by a Stanford Research Systems CS580 voltage-controlled current source. In the second, a Keithley 2182A nanovoltmeter was operated together with a Keithley 6221 current source in delta mode. $E$-$J$ and $\rho$-$J$ characteristics were measured using the pulse-delta mode of the 6221/2182A combination. The pulse width was 1 ms, the voltage was read 0.5 ms after the onset of each pulse, and consecutive pulses were separated by 200 ms.

\bibliographystyle{apsrev4-2}
\bibliography{jcpaper}

\section{Acknowledgments}

This work was supported by the DOE Office of Basic Energy Sciences, Materials Sciences and Engineering Division project 'Quantum fluctuations in narrow band systems'. C.G. acknowledges financial support by the European Research Council (ERC) under grant Free-Kagome (Grant Agreement No. 101164280). The high field experiments were performed at the National High Magnetic Field Laboratory, which is supported by National Science Foundation Cooperative Agreement No. DMR-2128556 and the State of Florida.

\section{Author contributions}

Crystals were synthesized and characterized by M. M. B., E. D. B., S. M. T., F. R., P. F. S. R. The experiment was designed by L. Z., C. G., P. J. W. M. The FIB microstructures were fabricated by L. Z., C. G., C. P. The magnetotransport measurements were performed by L. Z., C. G., D. G. The analysis of experimental results has been done by L. Z., C. G., P. J. W. M. All authors were involved in writing the paper.

\begin{figure*}
    \centering
    \includegraphics[width=1\linewidth]{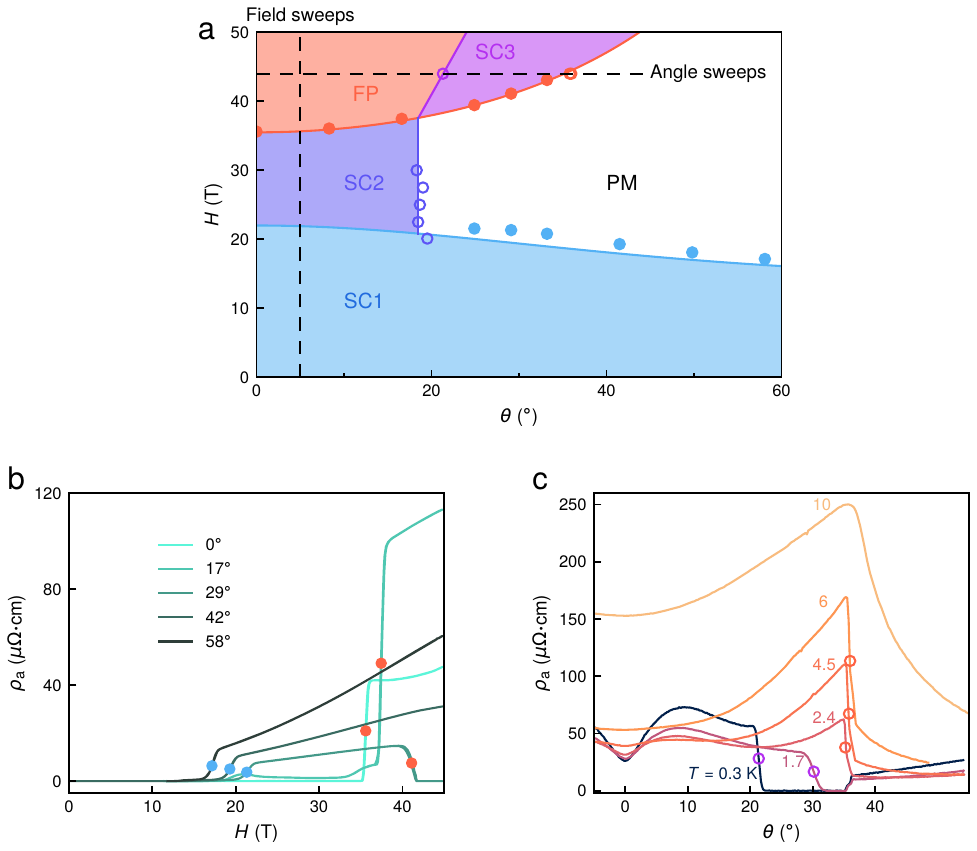}
   \caption{\textbf{Field--angle phase diagram of UTe$_2$ for fields in the $bc$ plane.}
\textbf{a} Phase diagram at $T = 0.3~\mathrm{K}$. Symbols mark phase boundaries
extracted from the measurements in \textbf{b} and \textbf{c}, with open
symbols from angle sweeps and filled symbols from field sweeps. Points on
the vertical SC2 phase boundary are taken from angle-sweep measurements in
our previous work\cite{zhangDimensionalityReinforcedSuperconductivity2025}.
\textbf{b} Field sweeps at various angles, $T = 0.3~\mathrm{K}$.
\textbf{c} Angle sweeps at at various temperatures, $H = 44~\mathrm{T}$.}
    \label{fig.1}
\end{figure*}

\begin{figure*}
    \centering
    \includegraphics[width=1\linewidth]{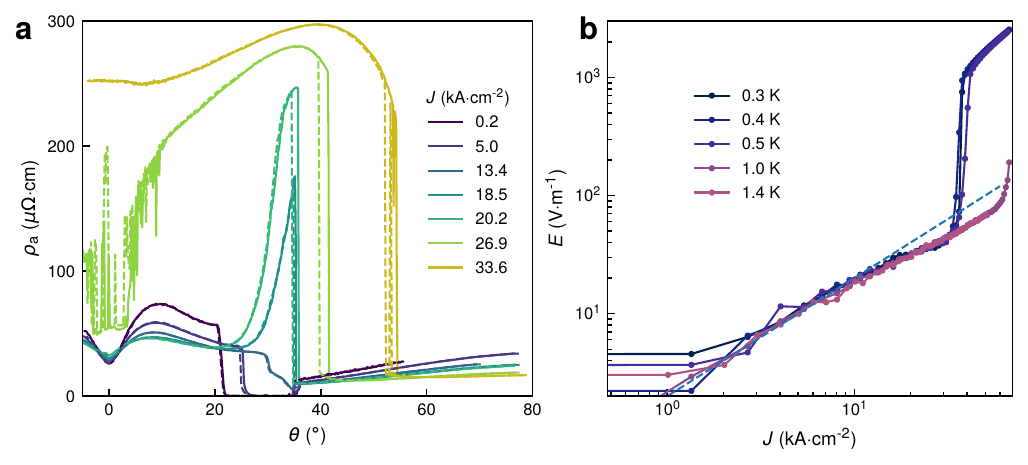}
    \caption{\textbf{a}. Resistivity measured as a function of angle at $T=0.3~\mathrm{K}$ for different DC current densities applied along the $a$ axis. Solid and dashed lines correspond to increasing and decreasing angle sweeps, respectively. \textbf{b}. $E$–$J$ characteristics plotted on logarithmic axes, measured at different temperatures at $H=44~\mathrm{T}$. The blue dashed line indicates a linear dependence, $E\propto J$.
}
    \label{fig.2}
\end{figure*}

\begin{figure*}
    \centering
    \includegraphics[width=1\linewidth]{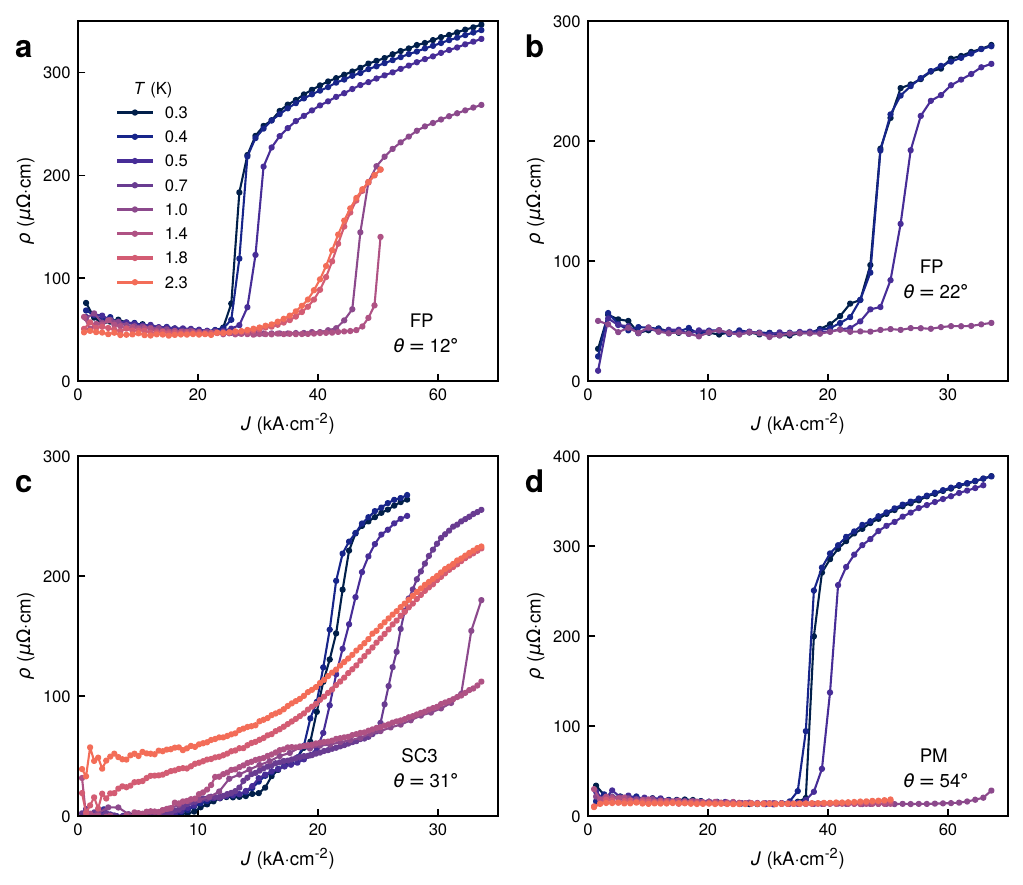}
    \caption{$\rho-J$ characteristics measured at 4 different angles at different temperatures at $H=44~\mathrm{T}$}
    \label{fig.3}
\end{figure*}

\begin{figure*}
    \centering
    \includegraphics[width=1\linewidth]{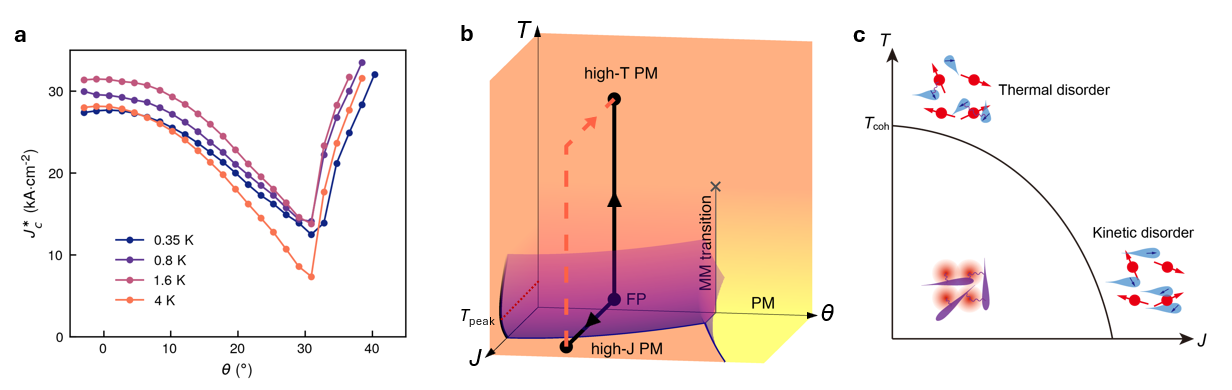}
    \caption{\textbf{a}. Extracted angle-dependent critical current density for the resistive transition at different temperatures at $H=44~\mathrm{T}$. \textbf{b}. Schematic phase diagram of the current-induced resistive transition (purple surface). \textbf{c}. Schematic phase diagram of a current-induced breakdown, illustrated for the case of a Kondo coherent state.}
    \label{fig.4a}
\end{figure*}

\end{document}


\renewcommand{\figurename}{Supplementary Figure}
\renewcommand{\tablename}{Supplementary Table}
\renewcommand{\thefigure}{\arabic{figure}}
\renewcommand{\thetable}{\arabic{table}}

\title{
{\large Supplementary Material for}\\[1em]
{\Large\bfseries Current-driven resistive switching in high-field heavy fermion metal UTe$_2$}
}
\author{Ling Zhang}
\email{ling.zhang@mpsd.mpg.de}
\affiliation{Max Planck Institute for the Structure and Dynamics of Matter, 22761 Hamburg, Germany}

\author{Chunyu (Mark) Guo}
\email{chunyu.guo@mpsd.mpg.de}
\affiliation{Max Planck Institute for the Structure and Dynamics of Matter, 22761 Hamburg, Germany}

\author{David Graf}
\affiliation{National High Magnetic Field Laboratory, Tallahassee, FL 32310, USA}

\author{Carsten Putzke}
\affiliation{Max Planck Institute for the Structure and Dynamics of Matter, 22761 Hamburg, Germany}

\author{Mitchell M. Bordelon}
\affiliation{Los Alamos National Laboratory, Los Alamos, NM 87545, USA}

\author{Eric D. Bauer}
\affiliation{Los Alamos National Laboratory, Los Alamos, NM 87545, USA}

\author{Sean M. Thomas}
\affiliation{Los Alamos National Laboratory, Los Alamos, NM 87545, USA}
\affiliation{Department of Physics, Colorado State University, Fort Collins, Colorado 80523, USA}

\author{Filip Ronning}
\affiliation{Los Alamos National Laboratory, Los Alamos, NM 87545, USA}

\author{Priscila F. S. Rosa}
\affiliation{Los Alamos National Laboratory, Los Alamos, NM 87545, USA}
\affiliation{Department of Physics, Colorado State University, Fort Collins, Colorado 80523, USA}

\author{Philip J. W. Moll}
\email{philip.moll@mpsd.mpg.de}
\affiliation{Max Planck Institute for the Structure and Dynamics of Matter, 22761 Hamburg, Germany}

\maketitle

\section{Reproducibility of the current-induced transitions }

\begin{figure}[htbp]
    \centering
    \includegraphics[width=0.75\linewidth]{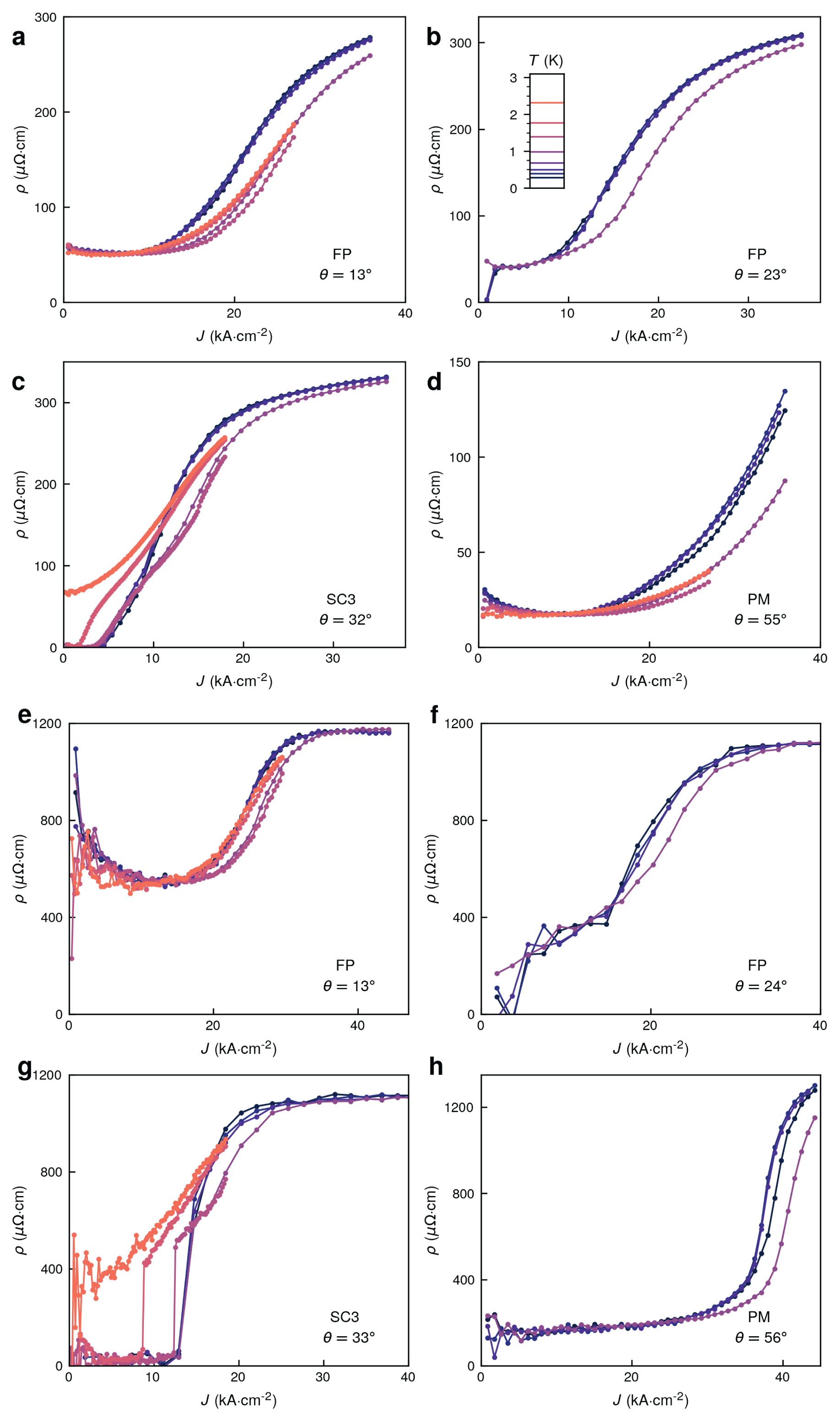}
    \caption{\textbf{a-d}. $\rho-J$ characteristics of sample S4 measured at 4 different angles at different temperatures at $H=44~\mathrm{T}$; \textbf{e-h}. $\rho-J$ characteristics of sample S5 measured at 4 different angles at different temperatures at $H=44~\mathrm{T}$.}
    \label{Sfig:rep}
\end{figure}

The current-induced transition into the high-resistivity state is observed in all four samples studied under high currents (S2--S5). The two additional samples (S4 and S5) included here differ from the two samples discussed in the main text in geometry and current direction ($J\parallel a$ and $J\parallel c$, respectively; see Supplementary Table~\ref{Stab:samples} and Supplementary Fig.~\ref{Sfig:rep}). Although the shape of the transition and the critical current density $J_c^*$ vary between samples (by up to $\pm 30\%$, depending on the angle), all samples exhibit the temperature dependence described in the main text: the transition shifts to higher current densities as the bath temperature is increased from $0.3\ \mathrm{K}$ to $1.6\ \mathrm{K}$, and, as in the main-text samples, this trend reverses at higher temperatures in S4 and S5.

The superconducting critical current density is likewise reproduced in the additional samples. Notably, for current applied along the $c$ axis, $j_c$ in the SC3 phase of sample S5 is approximately $10\ \mathrm{kA/cm^2}$ for $0.3\ \mathrm{K}<T<1.5\ \mathrm{K}$, comparable to the value obtained for $J\parallel a$. This indicates that $j_c$ in SC3 is largely isotropic with respect to the current direction, in clear contrast to the SC2 phase~\cite{zhangDimensionalityReinforcedSuperconductivity2025}, where $j_c$ along the $a$ axis exceeds that along the $c$ axis by at least a factor of 20.

\begin{table}[htbp]
  \begin{ruledtabular}
  \begin{tabular}{lccccl}
    Sample & Current direction & $L$ ($\mu$m) & $w$ ($\mu$m) & $t$ ($\mu$m) & Shown in \\
    \hline
    S1 & $a$ & 22.6& 2.5& 4.5& Fig.~1b \\
    S2 & $a$ & 4.9& 1.9& 1.6& Fig.~1c,2,3 \\
    S3 & $a$ & 10.9& 3.6& 1.2& Fig.~4a, Supplementary Fig.\ref{Sfig:jc_extraction}\\
    S4 & $a$ & 7.8& 1.8& 3.1& Supplementary Fig.\ref{Sfig:rep}\\
    S5 & $c$ & 3.2& 2.0& 1.4& Supplementary Fig.\ref{Sfig:rep}\\
    S6 & $a/c$ (meander)$^{*}$& 125.3& 2.2& 4.3& Supplementary Fig.\ref{Sfig:noise}\end{tabular}
  \end{ruledtabular}
  \begin{flushleft}
    \footnotesize $^{*}$Meander-shaped bar; current alternates between segments along $a$ and $c$.
  \end{flushleft}
  \caption{Summary of the samples measured in this work. $L$ denotes the distance between voltage contacts, and $w$ and $t$ the width and thickness of the current-carrying cross-section.}
  \label{Stab:samples}
\end{table}

\clearpage

\section{Bath-temperature dependence expected from thermal runaway}
\label{sec:thermal-runaway-T0-dependence}

In this section, we demonstrate that the anomalous temperature dependence of the critical current density $J_c^*$ cannot arise from ordinary thermal runaway. The intuitive fact that less heating power is needed to reach a given temperature when starting from a hotter sample is straightforward to show if the exact functional forms of the thermal and electrical conductivities are known. Such model-specific calculations, however, are subject to the caveat that deviations from the model assumptions might, in principle, still explain the data. Instead, we show, using only general thermodynamic principles, that higher bath temperatures necessarily imply lower critical currents under self-heating conditions. In particular, we place no restrictions on, and make no specific assumptions about, the heat-dissipation mechanism. Therefore, an increase in the critical current with increasing bath temperature cannot be explained by any thermal-runaway scenario. 
\begin{figure}
    \centering
    \includegraphics[width=0.75\linewidth]{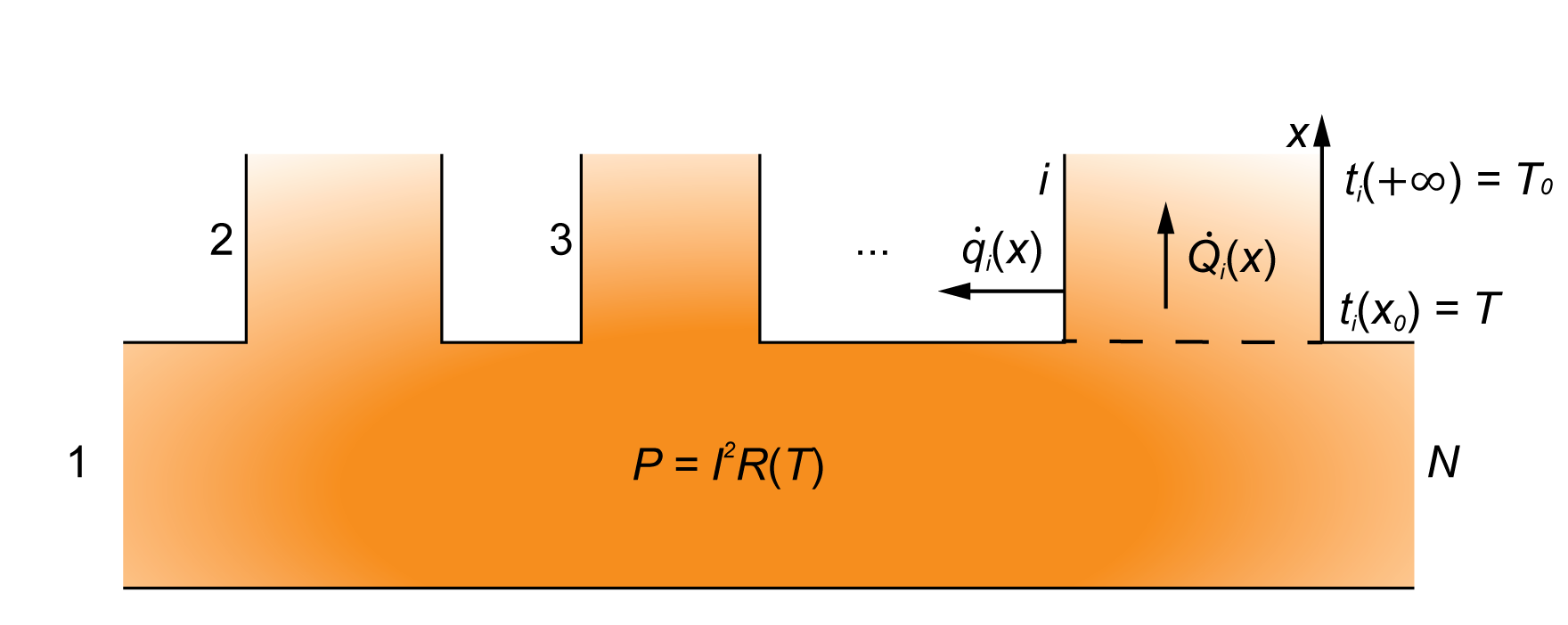}
    \caption{Schematic of the thermal-dissipation model. Joule heating $I^2R(T)$ in the resistive region is dissipated through multiple contacts. Along each contact $i$, heat is conducted longitudinally with heat current $\dot Q_i(x)$ while leaking into the surrounding environment at $T_0$ with heat flux $\dot q_i(x)$.}
    \label{Sfig:schematic}
\end{figure}

For a given bath temperature $T_0$, a steady state at sample temperature $T$ satisfies
\begin{equation}
    I^2 R(T) = \dot{Q}_{\mathrm{tot}}(T;T_0),
\end{equation}
where $R(T)$ is the resistance of the heated region and $\dot{Q}_{\mathrm{tot}}(T;T_0)$ is the total heat flow dissipated from the heated region to the bath. The largest current for which such a steady-state solution can exist is therefore
\begin{equation}
    I_c^2(T_0)
    =
    \max_{T>T_0}
    \frac{\dot{Q}_{\mathrm{tot}}(T;T_0)}{R(T)} .
    \label{eq:Ic_def_thermal_runaway}
\end{equation}
Here we assume that $R$ is determined by the local temperature of the resistive region, $R=R(T)$, while the bath temperature enters only through the thermal boundary condition. We also rely on positive thermal conductivity and positive heat-leakage conductance along all thermal paths.

We first consider the heat flow along the $i$th contact (Supplementary Fig.\ref{Sfig:schematic}. Let $t_i(x)$ be the local temperature profile along this contact, where $x=x_0$ denotes the boundary between the heated resistive region and the contact. The heat current carried along the contact is
\begin{equation}
    \dot{Q}_i(x)
    =
    -A_i \kappa_i[t_i(x)] \frac{d t_i}{dx},
    \label{eq:heat_current_contact}
\end{equation}
where $A_i$ is the cross-sectional area of the contact and $\kappa_i(t)$ is its thermal conductivity. With only the heat conduction along the contacts considered, $\dot{Q}_i(x)$ keeps constant along $x$. In realistic cases, heat can leak from the contact to the surrounding environment, for example through the substrate or the exchange gas/liquid. We describe this leakage phenomenologically by the heat flow per unit area
\begin{equation}
    \dot{q}_i[t_i(x),T_0]
    =
    \int_{T_0}^{t_i(x)} \frac{du}{r_i(u)},
    \label{eq:heat_leakage_density}
\end{equation}
where $r_i(u)>0$ is the effective temperature-dependent thermal resistance per unit area for the leakage channel. This form allows for nonlinear temperature dependence of the heat leakage.

Energy conservation over an infinitesimal segment of the contact gives
\begin{equation}
    \frac{d}{dx}
    \left[
        -A_i \kappa_i(t_i) \frac{dt_i}{dx}
    \right]
    =
    -W_i
    \int_{T_0}^{t_i(x)} \frac{du}{r_i(u)},
    \label{eq:heat_equation_contact}
\end{equation}
where $W_i$ is the width of the contact. We define
\begin{equation}
    f_i(t)=\int^t \kappa_i(u)\,du,
\end{equation}
so that $df_i/dx=\kappa_i(t_i)dt_i/dx$. Equation~\eqref{eq:heat_equation_contact} can then be written as
\begin{equation}
    \frac{d^2 f_i}{dx^2}
    =
    \frac{W_i}{A_i}
    \int_{T_0}^{t_i(x)} \frac{du}{r_i(u)} .
    \label{eq:f_equation_contact}
\end{equation}
Multiplying both sides by $df_i/dx$ and using the boundary conditions
\begin{equation}
    t_i(x\rightarrow\infty)=T_0,
    \qquad
    \frac{df_i}{dx}\bigg|_{x\rightarrow\infty}=0,
\end{equation}
one obtains the first integral
\begin{equation}
    \left(\frac{df_i}{dx}\right)^2
    =
    2\int_{T_0}^{t_i(x)}
    \frac{W_i}{A_i}
    \left(
        \int_{T_0}^{t'} \frac{du}{r_i(u)}
    \right)
    \kappa_i(t')\,dt' .
    \label{eq:first_integral_contact}
\end{equation}
Since the temperature decreases away from the heated region, $dt_i/dx<0$ and hence $df_i/dx<0$. The heat current flowing from the heated region into the $i$th contact is therefore
\begin{equation}
    \dot{Q}_i(T,T_0)
    =
    A_i
    \left[
    2\int_{T_0}^{T}
    \frac{W_i}{A_i}
    \left(
        \int_{T_0}^{t'} \frac{du}{r_i(u)}
    \right)
    \kappa_i(t')\,dt'
    \right]^{1/2},
    \label{eq:Qi_general}
\end{equation}
where $T=t_i(x_0)$ is the temperature at the center of the sample. The total dissipated heat flow is
\begin{equation}
    \dot{Q}_{\mathrm{tot}}(T,T_0)
    =
    \sum_i \dot{Q}_i(T,T_0).
    \label{eq:Qtot_general}
\end{equation}

We now examine the bath-temperature dependence of Eq.~\eqref{eq:Qi_general}. Define
\begin{equation}
    F_i(T,T_0)
    =
    2\int_{T_0}^{T}
    \frac{W_i}{A_i}
    \left(
        \int_{T_0}^{t'} \frac{du}{r_i(u)}
    \right)
    \kappa_i(t')\,dt' ,
\end{equation}
so that $\dot{Q}_i(T,T_0)=A_i\sqrt{F_i(T,T_0)}$. Because $\kappa_i(t)>0$ and $r_i(t)>0$, increasing $T_0$ reduces both the integration range and the inner leakage integral. More explicitly,
\begin{equation}
    \frac{\partial F_i}{\partial T_0}
    =
    -2\frac{W_i}{A_i}
    \frac{1}{r_i(T_0)}
    \int_{T_0}^{T}\kappa_i(t')\,dt'
    <0 .
    \label{eq:Fi_monotonic}
\end{equation}
Therefore, for fixed $T$,
\begin{equation}
    \frac{\partial \dot{Q}_i(T,T_0)}{\partial T_0}<0,
    \qquad
    \frac{\partial \dot{Q}_{\mathrm{tot}}(T,T_0)}{\partial T_0}<0 .
    \label{eq:Qtot_monotonic}
\end{equation}

It follows that, for two bath temperatures $T_1>T_2$,
\begin{equation}
    \dot{Q}_{\mathrm{tot}}(T;T_1)
    <
    \dot{Q}_{\mathrm{tot}}(T;T_2)
\end{equation}
at fixed $T$. In addition, the maximization domain $T>T_1$ is a subset of the domain $T>T_2$. Using Eq.~\eqref{eq:Ic_def_thermal_runaway}, we obtain
\begin{equation}
\begin{aligned}
    I_c^2(T_1)
    &=
    \max_{T>T_1}
    \frac{\dot{Q}_{\mathrm{tot}}(T;T_1)}{R(T)} \\
    &\le
    \max_{T>T_2}
    \frac{\dot{Q}_{\mathrm{tot}}(T;T_2)}{R(T)} \\
    &=
    I_c^2(T_2).
\end{aligned}
\label{eq:Ic_monotonic}
\end{equation}
Thus, within this general thermal-runaway model,
\begin{equation}
    T_1>T_2
    \quad\Rightarrow\quad
    I_c(T_1)\le I_c(T_2).
\end{equation}
That is, the thermal-runaway critical current is a non-increasing function of the bath temperature. Under ordinary conditions, it decreases with increasing $T_0$.

This result is opposite to the experimentally observed initial increase of the critical current with increasing temperature. The anomalous temperature dependence of $J_c^*$ therefore cannot be reproduced by any conventional thermal-runaway mechanism in which increasing $T_0$ only reduces the available thermal headroom for heat dissipation. This supports the conclusion that the observed critical current reflects an intrinsic current-driven instability rather than ordinary Joule-heating runaway.

\section{Quantifying the vanishing resistivity in SC3}

Since the discovery of the SC3 phase in UTe$_2$ by transport measurements\cite{ranExtremeMagneticFieldboosted2019}, its superconducting nature has been supported by contactless radio-frequency measurements\cite{lewinHighfieldSuperconductingHalo2025, wuQuantumCriticalLine2025} and magnetocaloric signatures consistent with bulk superconductivity~\cite{schonemannSuddenAdiabaticitySignals2023}, although direct thermodynamic evidence remains scarce. Here we provide an additional constraint by quantifying how closely the resistivity in SC3 approches zero.

\begin{figure}[t]
    \centering
    \includegraphics[width=0.75\linewidth]{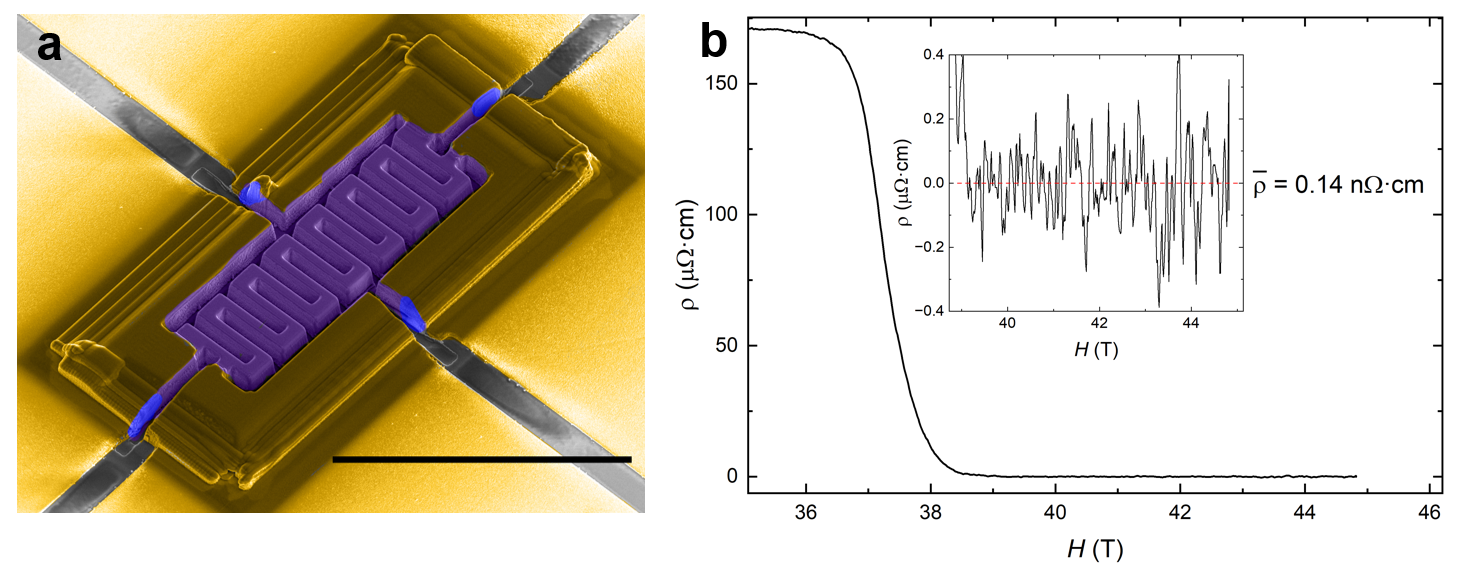}
    \caption{\textbf{a.} SEM image of meander-shape sample S6. Scale bar, $50~\mathrm{\mu m}$.  \textbf{b.} Field dependence of the resistivity of sample S6 at $\theta = 24^\circ$ (from $b$ towards $c$) and $T = 0.3~\mathrm{K}$. The resistivity is evaluated using the total length of the $c$-axis segments, $l_{\parallel c}$, and therefore represents an upper estimate (see text). Inset: enlarged view of the noise background within SC3. The red dashed line marks the averaged value $\bar{\rho}_{\mathrm{SC3}} = 0.14~\mathrm{n\Omega\,cm}$.}
    \label{Sfig:noise}
\end{figure}

For this estimate we use sample S6 (Supplementary Fig.\ref{Sfig:noise}a), whose meander shape provides a particularly long current path. The measured section consists of 12 segments along $c$ with a total length $l_{\parallel c} = 12\times 10.44 = 125.28~\mu\mathrm{m}$, and 13 segments along $a$ with a total length $l_{\parallel a} = 13\times 2.205 = 28.67~\mu\mathrm{m}$. Although the current distribution at the corners is difficult to model, the $a$ segments and the corners can only add to the dissipative path, so that the effective length satisfies $l_{\mathrm{eff}} > l_{\parallel c}$. Using $l_{\parallel c}$ therefore yields a conservative upper estimate of the resistivity, $\rho = RS/l_{\parallel c} > RS/l_{\mathrm{eff}}$, where $S$ is the cross-section of the segments.

Supplementary Fig.~\ref{Sfig:noise}b shows a field sweep at $\theta = 24^\circ$ and $T = 0.3~\mathrm{K}$, where SC3 sets in at $H \approx 38~\mathrm{T}$. Averaging the noise background in the window $H = 40$--$45~\mathrm{T}$ gives $\bar{\rho}_{\mathrm{SC3}} = 0.14~\mathrm{n\Omega\,cm}$, consistent with zero within the statistical uncertainty of the average. In the same sweep, the resistivity of the paramagnetic (PM) normal state at $H = 35~\mathrm{T}$, just below the metamagnetic (MM) transition into SC3, is $\rho_{\mathrm{PM}} = 170~\mu\Omega\,\mathrm{cm}$, evaluated with the same geometric convention. The ratio $\rho_{\mathrm{PM}}/\bar{\rho}_{\mathrm{SC3}} \approx 1.2\times 10^{6}$ is therefore independent of the uncertainty in $l_{\mathrm{eff}}$. A reduction of six orders of magnitude cannot arise in a metal from an increase in carrier density or a decrease in scattering rate, because the residual resistivity is limited by impurity and defect scattering, which is unchanged across the MM transition. The low value of resistivity is also more than an order of magnitude below the low-temperature residual resistivity of high-purity annealed copper\cite{matulaElectricalResistivityCopper1979} ($\sim$\,2~n$\Omega$\,cm), even without accounting for the sizeable magnetoresistance copper would exhibit at these fields. This further supports the superconducting nature of the SC3 phase.

\section{Extraction of the critical current density for the transition into the high-resistivity state}

We extract the critical current density $J_c^*$ of the current-induced resistive transition using a threshold criterion. In contrast to the superconducting transition, where a finite voltage emerges from a zero-resistance background, the transition here occurs on top of a finite resistive background whose magnitude depends on both field angle and temperature. We therefore choose fixed threshold resistivities $\rho_{\mathrm{th}}$ that lie within the transition region for most angles and temperatures. Below we show that, although the absolute value of $J_c^*$ depends on the choice of $\rho_{\mathrm{th}}$, its angular and temperature dependences are robust against this choice.

\begin{figure}
    \centering
    \includegraphics[width=0.8\linewidth]{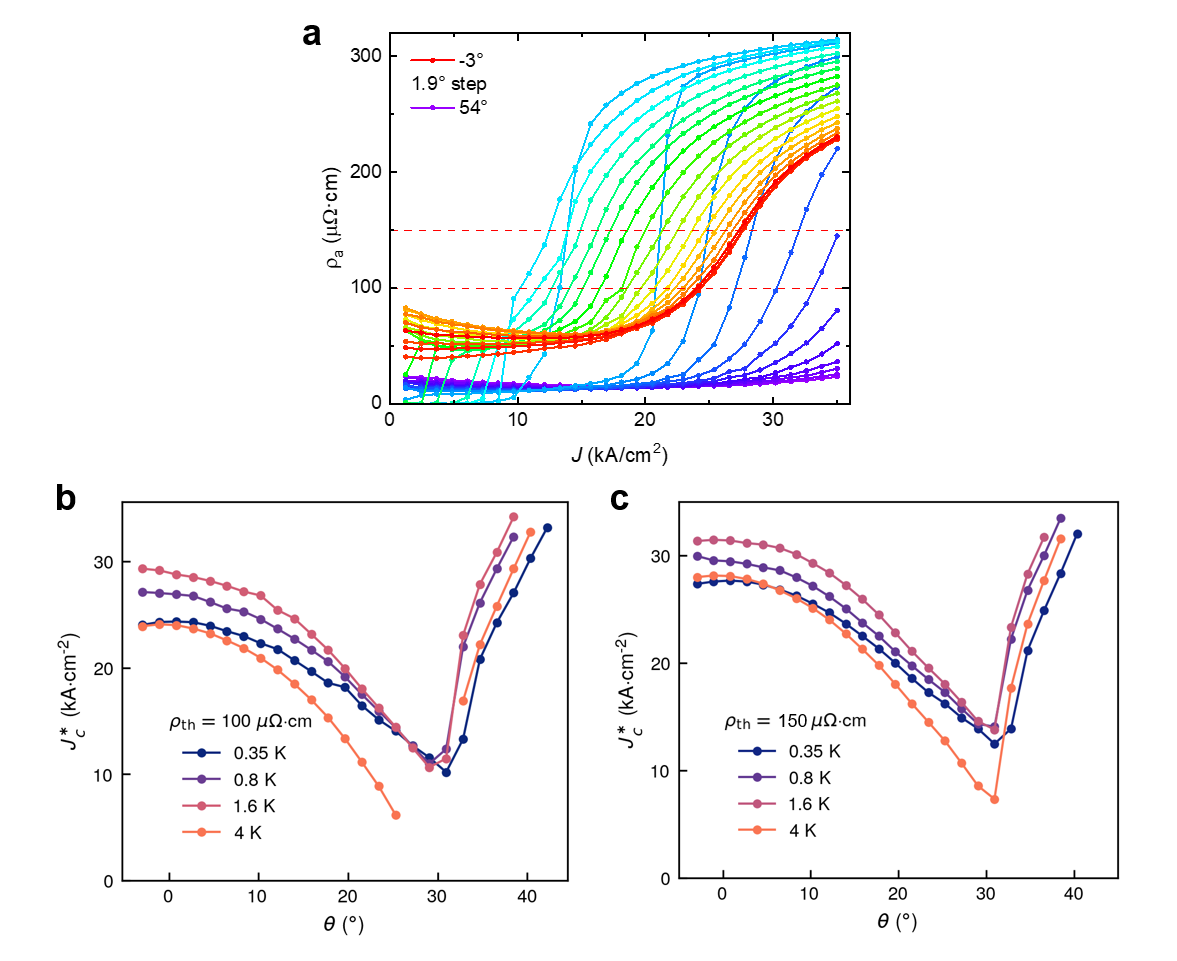}
    \caption{Threshold dependence of the extracted critical current density $J_c^*$. \textbf{a.} $\rho-J$ characteristics measured at $H=44~\mathrm{T}, T=0.3~\mathrm{K}$, for field angles in steps of $1.9^\circ$, the red dashed lines denote the resistivity thresholds. \textbf{b-c.} Temperature and angle dependence of $J_c^*$ extracted using \textbf{b.} $\rho_{\mathrm{th}}=100~\mathrm{\mu \Omega\cdot cm}$, \textbf{c.} $\rho_{\mathrm{th}}=150~\mathrm{\mu \Omega\cdot cm}$}
    \label{Sfig:jc_extraction}
\end{figure}

Figure~\ref{Sfig:jc_extraction}a shows $\rho$--$J$ curves measured on sample S3 at $T=0.3~\mathrm{K}$ for field angles in steps of $1.9^\circ$. The red dashed lines mark the thresholds $\rho_{\mathrm{th}}=150~\mathrm{\mu \Omega\cdot cm}$, used in Fig.~4a of the main text, and $\rho_{\mathrm{th}}=100~\mathrm{\mu \Omega\cdot cm}$, shown for comparison. Both lie above the background resistivity before the transition and, for most angles, below the resistivity reached after the transition. The values of $J_c^*$ extracted with $\rho_{\mathrm{th}}=100~\mathrm{\mu \Omega\cdot cm}$ are shown in Fig.~\ref{Sfig:jc_extraction}b for comparison with Fig.~\ref{Sfig:jc_extraction}c, which reproduces Fig.~4a of the main text. For both thresholds, $J_c^*$ exhibits a dip at the metamagnetic (MM) transition, and for most angles it reaches its maximum at $T=1.6~\mathrm{K}$. The only exception occurs near the MM transition for $\rho_{\mathrm{th}}=100~\mathrm{\mu \Omega\cdot cm}$. This deviation arises because the lower threshold is close to the resistivity reached after the superconducting depinning transition, so the extracted $J_c^*$ is partly affected by the superconducting transition rather than reflecting the high-resistivity transition alone. Apart from this exception, the angular and temperature dependences of $J_c^*$ are insensitive to the choice of $\rho_{\mathrm{th}}$.

\section{Estimate of the electric-field scale for suppressing Kondo coherence}

Here we provide an order-of-magnitude estimate of the electric field required to substantially perturb Kondo coherence in a metallic Kondo lattice. We adopt a simplified phenomenological picture in which a dc electric field shifts the quasiparticle distribution. Within the relaxation-time approximation, the characteristic displacement in momentum space is
\begin{equation}
    \delta k \simeq \frac{eE\tau_{\mathrm{tr}}}{\hbar},
\end{equation}
where \(e\) is the elementary charge and \(\tau_{\mathrm{tr}}\) is the transport relaxation time. The corresponding energy scale associated with the distortion of the quasiparticle distribution is
\begin{equation}
    \delta\varepsilon
    \sim \hbar v_F \delta k
    \sim eE v_F\tau_{\mathrm{tr}}
    = eE l_{\mathrm{tr}},
\end{equation}
where \(v_F\) is the Fermi velocity and \(l_{\mathrm{tr}}=v_F\tau_{\mathrm{tr}}\) is the transport mean free path. Physically, \(eE l_{\mathrm{tr}}\) represents the characteristic energy gained from the electric field over a transport mean free path.

As a phenomenological criterion for a substantial perturbation of the coherent Kondo-lattice state, we compare this energy with the lattice-coherence scale:
\begin{equation}
    eE_c l_{\mathrm{tr}} \sim k_B T_{\mathrm{coh}}.
\end{equation}
This gives the characteristic electric-field scale
\begin{equation}
    E_c \sim \frac{k_B T_{\mathrm{coh}}}{e l_{\mathrm{tr}}}.
    \label{eq:Kondo_decoherence_field}
\end{equation}
This criterion should be regarded as an order-of-magnitude estimate rather than a microscopic threshold for Kondo breakdown.

For UTe$_2$, we take \(T_{\mathrm{coh}}\sim 10~\mathrm{K}\) as estimated from STM measurements\cite{jiaoChiralSuperconductivityHeavyfermion2020}. Quantum-oscillation measurements provide a lower-bound estimate of approximately \(l_{\mathrm{q}}\sim 200~\mathrm{nm}\) for the quantum mean free path\cite{eatonQuasi2DFermiSurface2024}. The transport mean free path \(l_{\mathrm{tr}}\) is generally expected to be at least comparable to, and potentially longer than, \(l_{\mathrm{q}}\), particularly when small-angle scattering is important. Taking \(l_{\mathrm{tr}}\gtrsim l_{\mathrm{q}}\) therefore gives
\begin{align}
    E_c
    &\lesssim \frac{k_B T_{\mathrm{coh}}}{e l_{\mathrm{q}}} \nonumber\\
    &= \frac{
        \left(8.617\times10^{-5}~\mathrm{eV\,K^{-1}}\right)
        \left(10~\mathrm{K}\right)
    }{
        2\times10^{-7}~\mathrm{m}
    } \nonumber\\
    &\simeq 4.3\times10^{3}~\mathrm{V\,m^{-1}}.
\end{align}
Thus, within this simplified relaxation-time estimate, the characteristic field for strongly perturbing Kondo coherence is \(E_c\lesssim 4\times10^{3}~\mathrm{V\,m^{-1}}\). Because the relation between \(l_{\mathrm{tr}}\) and \(l_{\mathrm{q}}\), as well as the microscopic response of the Kondo lattice to a driven distribution, is not known precisely, this value should be interpreted as an approximate upper estimate within the assumptions above.

\bibliographystyle{apsrev4-2}
\bibliography{jc paper}